# Theoretical specific heat, and thermal conductivity estimated by detailed phonon vibrations

Hirofumi Kakemoto[1*]

[1]Clean Energy Research Center, University of Yamanashi, 4-3-11, Takeda, Kofu, Yamanashi 400-8511, Japan

We report the calculation results of specific heat ($c_v$), and thermal conductivity ($\kappa$) by using Einstein, and Debye models about rock salt (NaCl), oxides: $Na_xCoO_2$, $SrTiO_3$, and $LiNbO_3$. In calculation, the longitudinal (L), and transverse (T) sound velocities ($\nu_T$, $\nu_L$) were estimated from acoustic phonons' dispersions ($\sim \omega/K$) of above materials, and the average sound velocities ($\nu_a$) were input into Debye model for $\kappa$ equation, and results were compared with that of Einstein model. In some oxides, $\nu_a$ is relatively reduced at slight high $\nu_T$ ($\nu_T/\nu_L$=0.3-0.5). The relation of $\kappa$, $\nu_a$ and $T$ were plotted as the contour images about realizing low $\kappa$ value to application of thermoelectric properties.

(Dated: 1 July 2018 )

## 1. Introduction

Nowadays, various thermoelectric (TE) materials are reported about their variety of thermal transport properties of metallic ($Bi_2Te_3$) [1], semiconducting ($Si_{1-x}Ge_x$), and oxides. Recently, oxide TE materials are showing the favorable non-dimensional figure of merit ($ZT$). Particularly, $Na_xCoO_2$ (NCO) shows $p$-type, reported as $ZT$=0.8 at 800K [2], and $SrTiO_3$ (STO) shows $n$-type, and reported as $ZT$=0.3 at 1100K.[3] The favorable $n$-type TE oxide materials are demanded to develop for realizing high performance $p$-$n$ pairs of TE module use.[1-5] In theoretical thermal dynamics, specific heat ($c_v$) is well known by Einstein and Debye models. At first, Einstein represented about equivalent gas kinetic equation about thermal conductivity ($\kappa$), as $\kappa=(1/3)c_v\nu l$, where $l$ is mean free path, and $\nu$ is sound velocity.[6] Einstein developed for $c_v$ model. Debye represented for $c_v$, and $\kappa$ donated as $\kappa_{min}$.[7-9] Heat capacity ($C=mc_v$) is transferred to $\kappa$ by $\kappa=(1/V) \int D(\omega)C(\omega)\nu(\omega)l(\omega)d\omega$, where $V$, and $D$ are volume, and density, respectively, [10] and $\kappa$ is represented in eq.(2,4). $\kappa$ of nano-ordered grain and its boundaries is discussed.[11,12] In silicon clusters, weak phonon vibrations ("rattling") for lowering $\kappa$ is accounted by Einstein and/or Debye models.[13] In material, phonon vibrations' modes are important for thermal transport properties: for example, low $\kappa$ in $Bi_2Te_3$ caused because of averaged $\nu$ and heavy-Te phonon motions. $\kappa$ in $Na_xCoO_2$ also shows relatively low value because of $\nu_a$. On the other hand, $\kappa$ in $SrTiO_3$ is slight high value because of symmetric light-Ti phonon motions.

In this paper, calculation results about $c_v$ and $\kappa$ by using Einstein and Debye models about rock solute (NaCl), NCO, STO, and $LiNbO_3$ (LNO) are reported.

## 2. Calculation

The specific heat ($c_v$) and thermal conductivity ($\kappa$) representations are known by Einstein, and Debye models.[8] In Einstein model, $c_v$ and $\kappa$ cannot be affected for phonon properties of materials because of calculated by the Einstein frequency ($\omega_E$). Eventually, $c_v$ and $\kappa$ are not shown correctly at low temperature.

$$c_v^{Ein}=3nk_B(\theta_E/T)^2 e^{(\theta_E/T)}/(e^{(\theta_E/T)}-1)^2, \quad (1)$$
$$\kappa^{Ein}=(k_B^2/\pi\hbar)\, n^{1/3}\theta_E x^2 e^x/(e^x-1)^2, \quad (2)$$

where $n$ is the number of density of atoms, $x=h\nu/k_BT$, and $\theta_E$ is the Einstein temperature.

To select not $\omega_E$ but phonon frequencies of materials, above mentioned Einstein's result is possible to be modified Debye model. The average sound velocity ($\nu_a$) of longitudinal ($\nu_L$) and transverse ($\nu_T$) modes is written as $\nu_a=[(1/3)(1/\nu_L^3+2/\nu_T^3)]^{-1/3}$, and $c_v$ and $\kappa$ of Debye model are indicated as

$$c_v^{Debye}=9nk_B(T/\theta_D)^3 \int^{x_D} x^4 e^x\, dx/(e^x-1)^2, \quad (3)$$
$$\kappa^{Debye}=(\pi/6)^{1/3}k_B\, n^{2/3}\sum \nu_i (T/\theta_D)^2 \int x^3 e^x\, dx/(e^x-1)^2, \quad (4)$$

where $x=h\nu/k_BT$, and $\theta_D$ is the Debye temperature ($\theta_D=(h\nu_i/k_B)(6\pi^2 n)^{1/3}$).[8]

Table I lists $n$, sound of velocity ($\nu_L$, $\nu_T$, $\nu_a$), $\nu_p$, $\gamma$, $E$, $\theta_E$, and $\theta_D$ of materials.[14]

## 3. Result and discussion

Figure 1 shows sound (phonon) velocity of selected materials as listed in Table I. The average velocity ($v_a$) is calculated from longitudinal velocity ($v_L=\omega_{LA}/K$), transverse velocity ($v_T=\omega_{TA}/K$), and Einstein frequency ($\omega_E \simeq 3\times 10^{13}$ rad/s at 300K), and Debye frequency ($\omega_D$, $\omega_D^3 = 6\pi^2 Nv/V$) are depicted in Figs.1. In Einstein model, 3N harmonic oscillators are considered. The Einstein temperature ($\theta_E$) is an adjustable parameter in eq.(1,2), and frequency ($\omega_E$) is constant ($\omega_E > \omega_D$). In Debye model, 3N-6 harmonic oscillator is considered. In Fig.1(a), phonon dispersion of rock solute: NaCl, and $v_a$ is estimated to be 2,327 m/s from $v_L$ and $v_T$ by using $\omega/K$.[15,16] In Fig.1(b), $v_a$ of NCO is calculated to be 2,113 m/s from $v_L$ and $v_T$ by using $2\pi v/K$.[17] In Fig.1(c), $v_a$ of STO is also calculated to be 5,270 m/s from $v_L$ and $v_T$ by using $2\pi v/K$.[18] In Fig.1(d), $v_a$ of LNO is calculated to be 3,558 m/s.[19] From above results, Poisson's ratio ($v_p$), Grüneisen parameter ($\gamma$), and Young modulus ($E$) are estimated, and listed in Table I.[14]

$\kappa$ and $c_v$ are calculated by using eq.(1-4) with inputting $n$, $v_a$, and $\theta_E$, or $\theta_D$ in Table I.

Figure 2 shows $\kappa$ and $c_v$ as the functions of $v_a$, and $T$. In Fig.2(a), $\kappa^{Ein}$, $\kappa^{Debye}$, $c_v^{Ein}$, and $c_v^{Debye}$ of NaCl are showed to be 5.6 W/mK, 4.6 W/mK, 49.9 J/molK, and 45.4 J/molK at 1000K, respectively.[20] Although high $\kappa$ of Na (83.6 W/mK) caused by Na ion displacement is known, here low $\kappa$ is estimated for NaCl caused by phonon vibration of $v_a$. In Fig.2(b), $\kappa^{Ein}$, $\kappa^{Debye}$, $c_v^{Ein}$, and $c_v^{Debye}$ are also calculated for NCO to be 4.4 W/mK, 3.8 W/mK, 96.6 J/molK, and 89.4 J/molK at 1000K, respectively.[17] In Fig.2(c), $\kappa^{Ein}$, $\kappa^{Debye}$, $c_v^{Ein}$, and $c_v^{Debye}$ are estimated for STO to be 11.9 W/mK, 10.6 W/mK, 124 J/molK, and 122 J/molK at 1000K, respectively.[3,18] In Fig.2(d), $\kappa^{Ein}$, $\kappa^{Debye}$, $c_v^{Ein}$, and $c_v^{Debye}$ are calculated for LNO to be 5 W/mK, 4.3 W/mK, 98.5 J/molK, and 85 J/molK at 1000K, respectively.[21]

Estimated above results are correct for $c_v \sim 0$ at 0K, and Dulong-Petit value ($c_v/n = 3R = 24.4$ J/molK, $R$: gas constant) up to 1000K. In addition, $\kappa^{Debye}/\kappa^{Ein} = 0.82$-$0.89$, and $c_v^{Debye}/c_v^{Ein} = 0.83$-$0.98$ are estimated.

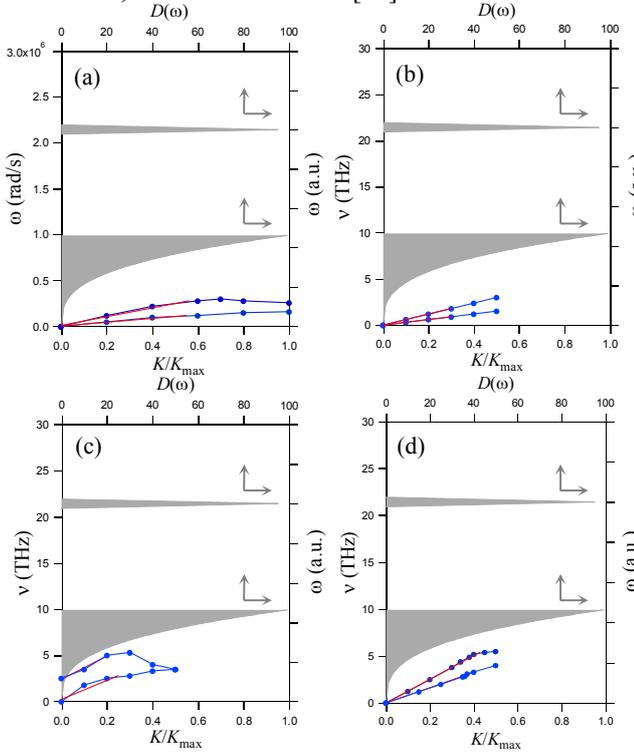

FIG. 1 The phonon dispersion: LA, TA versus wavenumber ($K$, $K_{max} \simeq 10^4$ cm$^{-1}$), (dots: reported, solid line: fitting), and density of the state of phonon versus Einstein, and Debye frequencies of (a) rock salt: NaCl, (b) Na$_x$CoO$_2$, (c) SrTiO$_3$, and (d) LiNbO$_3$.

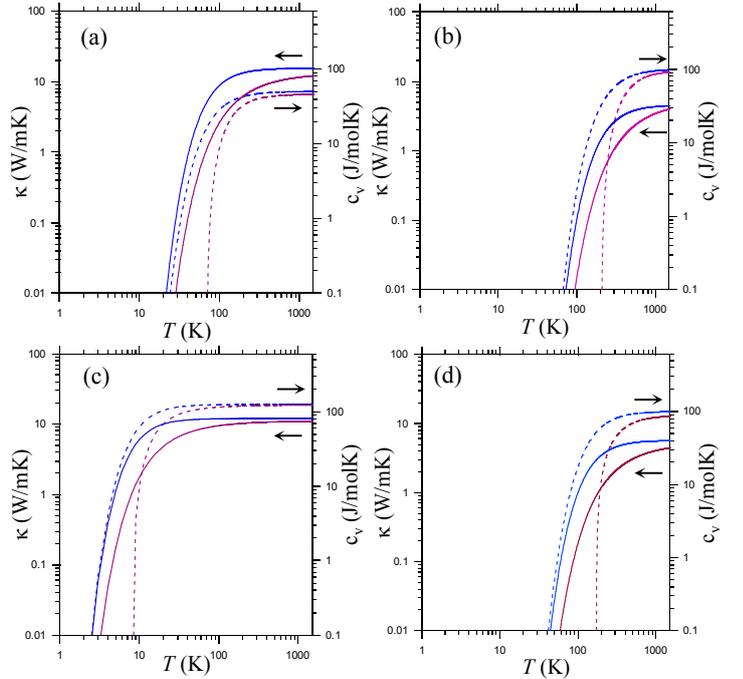

FIG. 2 Thermal conductivity (solid line) and specific heat (dashed line) calculated by Einstein ("blue"), and Debye models ("purple") for (a) rock salt: NaCl, (b) Na$_x$CoO$_2$, (c) SrTiO$_3$, and (d) LiNbO$_3$ as a function of $T$.

Table I Materials, carrier density, sound velocities, material constants, Einstein, and Debye temperature.

| Material | $N$ ($10^{19}$cm$^{-3}$) | $v_L$ | $v_T$ | $v_a$ ($10^2$cm/s) | $v_p$ | $\gamma$ | $E$ (GPa) | $\Theta_E$ (K) | $\Theta_D$ (K) |
|---|---|---|---|---|---|---|---|---|---|
| rock salt: NaCl | 4,460 | 5,500 | 2,050 | 2,327 | 0.42 | 2.87 | 36.1 | 270.3 | 265.8 |
| diamond | | 17,500 | 1,1640 | 12,729 | | | | | |
| Na | | | | 3,200 | | | | | |
| Na$_x$CoO$_2$ | 1,000 | 3,768 | 1,884 | 2,113 | 0.33 | 2.00 | | 820 | 793 |
| SrTiO$_3$ | < 1.0 | 7,850 | 4,770 | 5,270 | 0.21 | 1.31 | | 30 | 30 |
| LiNbO$_3$ | < 1.0 | 7,963 | 3,141 | 3,558 | 0.41 | 2.72 | 201 | 500 | 500 |

Figure 3 (a) shows $\nu_a$ versus $\nu_L$, and $\nu_T/\nu_L$, and Fig.3 (b) shows $\nu_a$ versus $\nu_T$, and $\nu_T/\nu_L$. As shown in Fig.3 (a), $\nu_a$ and $\nu_T/\nu_L$ are increased with increasing $\nu_L$. In Fig.3 (b), $\nu_a$ is increased with increasing $\nu_T$, but $\nu_L/\nu_T$ is decreased with increasing $\nu_T$. Here, $\nu_a$ is a key-factor for reducing $\kappa$. $\nu_a$ of NCO and LNO is relatively reduced by slight high $\nu_T$ ($\nu_T/\nu_L$=0.5 for NCO, =0.39 for LNO), as follows $\nu_a=[(1/3)(1/\nu_L^3+2/\nu_T^3)]^{-1/3}$. $\kappa_{min}$ of several material is usually estimated by Debye model using $\nu_a$ and $T$ in eq.(4). $\kappa$ should not be decided high displacement atom such as Na, but phonon dispersion in crystal, particularly, low $k$ is possible to be reduced by above mentioned $\nu_T$.

Figure 3 (c) shows the contour image of $\kappa$ ($\kappa_{min}$) as the functions of $\nu_a$ (1,500m/s<$\nu_L$<6,500m/s, $\nu_T$=5,000m/s) and $T$, and Fig.3 (d) also shows the contour image of $\kappa_{min}$ ($\theta_D$=30K) as the functions of $\nu_a$ (1,500m/s<$\nu_T$<6,500m/s, $\nu_L$=5,000m/s) and $T$. In Fig.3 (c), $\kappa_{min}$ shows high value in region of $\nu_a$=2,000-6,500m/s, on the other hand, in Fig.3(d), $\kappa_{min}$ is reduced in region of $\nu_a$=2,000-6,500m/s.

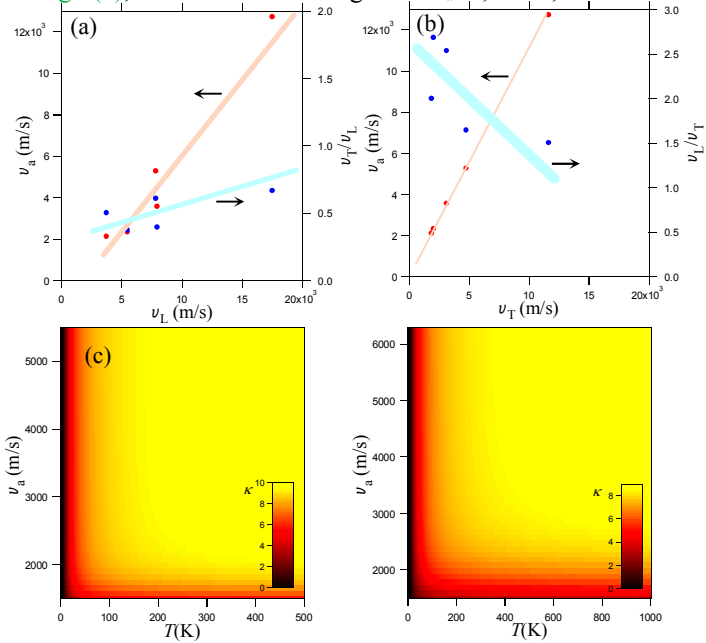

FIG. 3 The average sound velocity ($\nu_a$) versus (a) $\nu_L$, $\nu_T/\nu_L$, (b) $\nu_L$, $\nu_T/\nu_L$, (c) contour image about $\kappa_{min}$, (changed $\nu_L$, fixed $\nu_T$) and (d) contour image about $\kappa_{min}$, (changed $\nu_T$, fixed $\nu_L$) as the functions of $\nu_a$ and $T$.

**4. Conclusion**

The average sound of velocities ($\nu_a$) of TE materials are calculated by using sound velocities ($\nu_L$, $\nu_T$) from reported phonons' dispersions ($\sim\omega/K$). The thermal conductivity ($\kappa$) is possible to be estimated by $\kappa=(1/V) \int D(\omega)C(\omega)\nu(\omega)l(\omega)d\omega$ as follows Debye model. For TE material, particularly, $k$ and $c_v$ were compared Einstein and Debye models input by sound velocity of materials. $\nu_a$ is reduced by slight high $\nu_T$, and $\kappa$ donated as $\kappa_{min}$ is decreased with decreasing $\nu_a$ Estimated $\kappa_{min}$ can be considered as $\kappa_{ph}$ of $\kappa_{tot}=\kappa_{ph}+\kappa_e$.

In future study, the investigation of $\kappa$ about $n$-type Nb related TE oxide will be carried out by using Harman method (experiment) and Debye model (calculation).

Acknowledgment

This work was partly supported by Japan Society for the Promotion of Science (JSPS) KAKENHI Grant-in-Aid for Scientific Research(C) Number JP25410238.

*Present address: 1-15-11, Sakura-cho, Tsuchiura, Ibaraki, 300-0037, Japan, Techno Pro R&D company (Tsukuba branch), Techno Pro Inc.
e mail: hkkemoto@yamanashi.ac.jp